\documentclass[prl,twocolumn,superscriptaddress,showpacs,amsmath,amssymb]{revtex4-1}
\usepackage{graphicx,dcolumn,bm}

\begin{document}

\title{Construction of optimal witness for unknown two-qubit entanglement}

\author{H. S. Park}
\email[Corresponding author. ]{hspark@kriss.re.kr}
\affiliation{Korea Research Institute of Standards and Science, Daejeon 305-340, Korea}
\author{S.-S. B. Lee}
\affiliation{Department of Physics, Korea Advanced Institute of Science and Technology, Daejeon 305-701, Korea}
\author{H. Kim}
\author{S.-K. Choi}
\affiliation{Korea Research Institute of Standards and Science, Daejeon 305-340, Korea}
\author{H.-S. Sim}
\email[Corresponding author. ]{hssim@kaist.ac.kr}
\affiliation{Department of Physics, Korea Advanced Institute of Science and Technology, Daejeon 305-701, Korea}

\date{\today}

\begin{abstract}

Whether entanglement in a state can be detected, distilled, and quantified
without full state reconstruction is a fundamental open problem.
We demonstrate a new scheme encompassing these three tasks for arbitrary two-qubit entanglement,
by constructing the optimal entanglement witness for polarization-entangled mixed-state photon pairs
without full state reconstruction.
With better efficiency than quantum state tomography,
the entanglement is maximally distilled by newly developed tunable polarization filters,
and quantified by the expectation value of the witness, which equals the concurrence.
This scheme is extendible to multiqubit Greenberger-Horne-Zeilinger entanglement.

\end{abstract}

\pacs{03.65.Ud, 03.67.Mn, 42.65.Lm}

\maketitle

A fundamental issue is how to detect, distill, and quantify entanglement
in a state~\cite{Guhne_rev},
quantum correlation not imitated by classical correlation.
These tasks are essential in understanding
quantum nonlocality versus local realism~\cite{Bell},
and in developing quantum technologies where
entanglement needs to be enhanced and monitored.
An important open question is whether the tasks can be done
without referring to the full knowledge of the state~\cite{Plenio_sci}.
It is motivated from the demand of efficient protocols
that directly access relevant quantities.

Current schemes for those tasks rely on the full knowledge of the state estimated by
quantum state tomography (QST)~\cite{James},
even for two-qubit entanglement.
For example, the detection by an entanglement witness~\cite{Barbieri}, 
a Hermitian operator whose expectation value is positive for all separable states~\cite{Horodecki_witness,Terhal,Lewenstein},
requires QST to choose proper
measurement settings.
Procrustean distillation~\cite{Kwiat,Wang}, in which entanglement is enhanced by local filtering,
has been also performed with known entanglement or QST.
Since QST may have
redundant information and becomes impractical for systems with larger number of qubits~\cite{Plenio_sci},
those approaches are indirect and inefficient.
Hence, the schemes not relying on QST are necessary;
there is a quantification scheme~\cite{Walborn,Schmid},
which is however not always available as it requires two state copies.

\begin{figure}[bt]
\includegraphics[width=.5\textwidth]{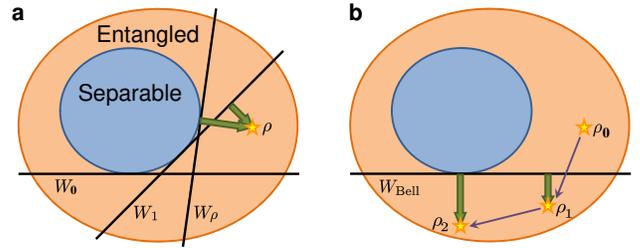}
\caption{(Color Online)
{(a)} Entanglement quantification by optimal witness $W_\rho$.
Each witness $W$ provides a supporting hyperplane $\mathrm{Tr}(W \rho') = 0$
(solid lines) of the set of separable states in the space (filled area) of two-qubit states.
The ``distance'' $d \equiv \max \{0, - 2 \mathrm{Tr}(W \rho) \}$ (thick arrows)
between a state $\rho$ and a hyperplane is maximized by $W_\rho$.
The maximum distance equals $\mathcal{C}(\rho)$.
{(b)} Construction of $W_\rho$ via Procrustean distillation.
The transformation from Bell witness $W_\textrm{Bell}$
to $W_\rho$ is
equivalent to that from $\rho$ ($=\rho_0$)
to the maximally distilled state $\rho_\textrm{dis}$ ($=\rho_{2}$),
the state optimally detected by $W_\textrm{Bell}$.
In our scheme, the distillation is achieved iteratively using tunable local filters;
the subscript
is the iteration index.
}
\label{witness}
\end{figure}

It has been found~\cite{Terhal_optimal,Brandao,Guhne_PRL,Eisert} that
entanglement detection and quantification can be done using
a physical observable known as optimal witness $W_\rho$.
$W_\rho$ is defined relative to a state $\rho$ as
$\mathrm{Tr}(W_\rho \rho) \equiv \min_{W \in \mathbb{M}} \mathrm{Tr} (W \rho)$,
where $\mathbb{M}$ is a collection of witnesses $W$
and the trace $\mathrm{Tr}(\cdot)$ gives the expectation value.
For a two-qubit state $\rho$, it gives
a popular entanglement measure,
concurrence $\mathcal{C} (\rho)$~\cite{Bennet_conc,Wootters_conc,Brandao,Verstraete_Thesis},
\begin{equation}
\mathcal{C} (\rho)
= \max \{0, -2 \mathrm{Tr} (W_\rho \rho) \}.
\label{witness_conc}
\end{equation}
The proper collection $\mathbb{M}$ here
contains all the witnesses which are obtained from
the Bell witness
$W_\mathrm{Bell}$ ($ = I / 2 - |\textrm{Bell} \rangle \langle \textrm{Bell} |$)
through stochastic local operations and classical communications (SLOCC),
a composition of local unitary operation and local filtering.
In Eq.~\eqref{witness_conc}, $\mathcal{C} (\rho)$ is interpreted as
the ``distance'' between $\rho$ and
the supporting hyperplane $\textrm{Tr}(W_\rho \rho')=0$ of
the set of separable states in the space of two-qubit states; see Fig.~\ref{witness}(a).
One can obtain $\mathcal{C}(\rho)$ by constructing $W_\rho$,
rather than indirectly from a calculation~\cite{Wootters_conc} based on QST estimation.

It has not been known how to construct $W_\rho$ without QST.
We will do it for two-qubit entanglement, based on
the following link between $W_\rho$ and Procrustean distillation~\cite{Kwiat,Wang};
see Fig.~\ref{witness}(b).
The optimal witness $W_{\rho_\textrm{dis}}$
relative to $\rho_\textrm{dis}$, the maximally distilled state obtained from $\rho$,
is constructed from $W_\textrm{Bell}$ by local unitary operation.
Moreover
$\textrm{Tr}(W_\rho \rho)$ and $\textrm{Tr}(W_{\rho_\textrm{dis}} \rho_\textrm{dis})$ are identical,
up to constant $s_0$; see below.
These facts, originated from the SLOCC invariance of $\mathcal{C}$, mean that
the distillation operation connecting $\rho$ and $\rho_\textrm{dis}$ is
equivalent to the filtering part of $W_\rho$.
We obtain $\rho_\textrm{dis}$ to construct $W_\rho$ below.

In this Letter, we provide a new entanglement detection scheme
not relying on QST,
based on $W_\rho$ and
individual copies of a state.
For a unknown two-qubit mixed state $\rho$ of polarization-entangled photon pairs,
we construct $W_\rho$ and $W_{\rho_\textrm{dis}}$
(hence determine $\mathcal{C}(\rho)$ and
$\mathcal{C}(\rho_\textrm{dis})$),
by maximally distilling $\rho$ and then
measuring
two-qubit correlation with applying local unitary operation.
This approach is more efficient than QST for the distillation and quantification,
and extendible to multiqubits.

\begin{figure}[bt]
\includegraphics[width=.42\textwidth]{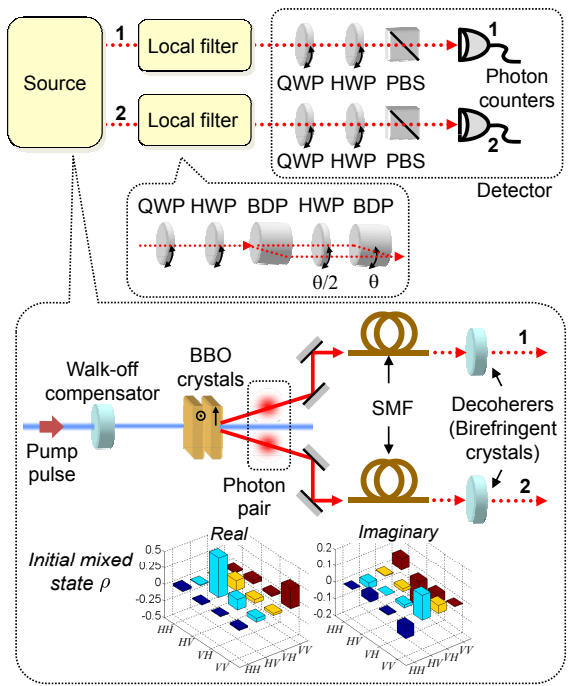}
\caption{(Color Online) {Setup.}
In the source, polarization-entangled photon pairs are generated
from spontaneous parametric down-conversion (SPDC)
in two cascaded beta-barium borate (BBO) crystals pumped by pre-walk-off-compensated pulses.
After spatial-mode filtering by single-mode fibers (SMFs),
decoherers
convert them into a mixed state $\rho$.
Then,
entanglement in $\rho$ is distilled by local filters,
in which a tunable polarization-dependent loss can be applied with an arbitrary basis
using beam-displacing prisms (BDPs).
Finally, filter-transmitted photon pairs are analyzed with quarter-wave plates (QWPs),
half-wave plates (HWPs),
polarizing beam splitters (PBSs), and SMF-coupled photon counters.
Our scheme never uses the QST estimation of $\rho$ (inset).
}
\label{scheme}
\end{figure}

We describe the experiment setup in Fig.~\ref{scheme}.
The source generates an unknown mixed state $\rho$
of polarization-entangled photon pairs,
in which
qubit $|0 \rangle$ ($|1 \rangle$) is encoded in horizontal $H$ (vertical $V$) photon
polarization.
Here, a pure Bell state of photon pairs, prepared through SPDC~\cite{Kwiat_SPDC,Nambu_SPDC},
becomes $\rho$
due to asymmetry of pump polarization and severe decoherence
applied in decoherers~\cite{Kwiat}
composed of birefringent quartz crystals~\cite{SUPP}.
Next, photon $j$ ($=1,2$) of $\rho$ moves to
local filter $j$.

\begin{figure}[bt]
\includegraphics[width=.42\textwidth]{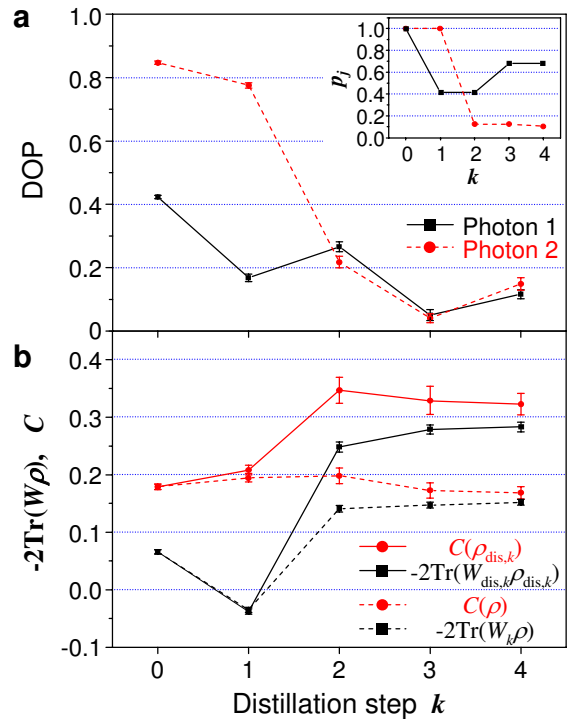}
\caption{(Color Online) { Distillation and Quantification.}
{(a)} Degree of polarization (DOP) and filter transmission probability $p_j$ (inset) at distillation step $k$.
At $k=0$, no filtering is applied.
{(b)} The values of $-2 \textrm{Tr} (W_k \rho)$,
$-2 \textrm{Tr} (W_{\textrm{dis},k} \rho_{\textrm{dis},k})$,
$\mathcal{C}(\rho)$, and $\mathcal{C}(\rho_{\textrm{dis},k})$;
$W_k$ and $W_{\textrm{dis},k}$ are
the best witnesses for $\rho$ and the distilled state $\rho_{\textrm{dis}, k}$ at step $k$,
respectively, and $\mathcal{C}$'s are estimated by QST.
Error bars represent statistical uncertainty [$\pm \hbox{(counts)}^{1/2}$].
Data points in (b) has the same size as the error bar
($\pm 0.004$) for $W$'s at $k=0$.
For larger $k$, the error bars for $W$'s increase due to smaller $N/M$.
}
\label{iteration}
\end{figure}

Our scheme requires that
any arbitrary filtering (SLOCC) operation is realized by the local filters.
For this purpose, we have newly designed the filters~\cite{HKim}.
By filter $j$, photons having polarization parallel to its filtering basis
are filtered off (not proceeding to detector $j$)
with probability $1-p_j$, while those with orthogonal polarization are always transmitted.
The filtering basis and transmission probability $p_j$ of filter $j$ are tunable
over the whole range
by a QWP and a HWP and by two BDPs and the HWP in between, respectively,
while preserving interferometric stability;
see the inset of Fig.~\ref{iteration}(a).

In the detectors, filter-transmitted photon pairs
undergo local unitary rotation of photon polarization,
realized with a QWP and a HWP, and
are projected onto
a two-qubit product state by PBSs.
Coincidence count $\langle n_1 n_2 \rangle$ is obtained
from photon number $n_j$ detected at counter $j$.
For our severely decohered state $\rho$,
we choose the measurement time of 10 s
(during which the source creates $5 \times 10^4$ photon pairs)
per detector setting.

We first maximally distill $\rho$ iteratively, based on the fact~\cite{Verstraete_normal}
that the degree of polarization (DOP) of each photon of $\rho_\textrm{dis}$
vanishes, i.e., all the local reduced density matrices
are fully mixed.
We
first (at step $k=1$) erase DOP 1 of photon 1 by tuning filter 1,
next (at $k=2$) do the same for photon 2,
and then (at $k=3,\cdots$)
repeatedly re-adjust
filters 1 and 2 alternately,
until both the DOPs vanish within experimental uncertainty~\cite{Lee}.
At each step for photon $j$,
the filtering basis and transmission probability $p_j$ for erasing
are determined from
the measurement of
the local reduced density matrix~\cite{SUPP}.
This iterative distillation finishes usually within a few steps
and does not require the full reconstruction of $\rho$.

After the distillation, $W_\rho$ is constructed
with local unitary
operation on photon polarization. 
We consider
the two-qubit correlation of
$(4 \langle n_1 n_2 \rangle - 2 \langle n_1 \rangle - 2 \langle n_2 \rangle
+ N)/N$, which has the information of two-photon Stokes parameters~\cite{SUPP}.
$\langle n_1\rangle$ ($\langle n_2 \rangle$) is the
number of photons detected at detector 1 (2)
and $N$ is the total number of photon pairs passing through the filters
among $M$ pairs injected from the source.
There appear three different pairs of local extremum values of the two-qubit correlation,
$\pm \lambda_{1,2,3}$ ($\lambda_1 \ge \lambda_2 \ge \lambda_3 \ge 0$),
if one measures it with varying local unitary operation (by tuning the detector QWPs and HWPs).
At the step of the maximal distillation,
\begin{equation}
\mathrm{Tr} (W_\rho \rho) = \frac{s_0}{4} ( 1 - \lambda_1 - \lambda_2 + q \lambda_3),
\label{witness_opt}
\end{equation}
where $s_0 = N / (M \sqrt{p_1 p_2})$
and the sign factor $q = -1$ for all entangled states~\cite{Lee}. 
Moreover,
$\mathrm{Tr} ({W_{\rho_\textrm{dis}} \rho_\textrm{dis}}) = ( 1 - \lambda_1 - \lambda_2 + q \lambda_3)/4
= \mathrm{Tr} (W_\rho \rho) / s_0$.
Hence, by measuring $\lambda$'s, 
one constructs both $W_\rho$ and $W_{\rho_\textrm{dis}}$
and determines $\mathcal{C}(\rho)$ and $\mathcal{C}(\rho_\textrm{dis}) = \mathcal{C}(\rho)/s_0$
using Eq.~\eqref{witness_conc}.
We measure
$\lambda$'s efficiently~\cite{Lee},
by first determining the detector settings
for $\lambda$'s and then measuring $\lambda$'s.
We perform 16 independent detector settings of coincidence counting
for the former,
and 12 settings (4 settings/$\lambda$ $\times$ three $\lambda$'s) for the latter.
The 12 settings for $\lambda$'s are determined from the 16-setting data
through the singular value decomposition of two-photon Stokes parameters~\cite{SUPP}.
Crude determination of the settings for $\lambda$'s is enough for
precise detection of $\lambda$'s, since $\lambda$'s are extremum values;
a small error ($\sim \delta$) of the setting determination causes
a much smaller error ($\sim \delta^2$) of the detection of $\lambda$'s.

Before discussing our experimental data, we introduce the concept of the best witness.
Often one can explore not the whole set $\mathbb{M}$ of witnesses
but only its subset $\mathbb{M}'$ due to certain limitation,
and $\mathbb{M}'$ may not include the optimal witness $W_\rho$.
In this case, we propose to use the best witness $W_{\rho, \mathbb{M}'}$ within $\mathbb{M}'$,
$\textrm{Tr} (W_{\rho, \mathbb{M}'} \rho) \equiv \textrm{min}_{W \in \mathbb{M}'} \textrm{Tr} (W \rho)$.
For two qubits,
it gives, via $\max \{0, -2 \mathrm{Tr} (W_{\rho, \mathbb{M}'} \rho) \}$, 
the best lower bound of $\mathcal{C} (\rho)$, which is the best among those achievable by
the witnesses in $\mathbb{M}'$.
In our experiment,
the witness constructed via Eq.~\eqref{witness_opt} is the best witness at the steps
where the maximal distillation is not achieved~\cite{SUPP}, and it becomes
the optimal witness at the step of the maximal distillation; see below.
This supports that our determination of $\mathcal{C}(\rho)$ does not exceed the exact value,
even if there is experimental imperfection.
And, it allows us to extend our scheme to the quantification of three-qubit
entanglement, for which it has not been known how to achieve $W_\rho$.
Note that the new theoretical findings of this work
include
the link between $W_\rho$ and the scheme in Ref.~\cite{Lee},
the best witness,
and the extension to multiqubits~\cite{SUPP}.

\begin{figure}[bt]
\includegraphics[width=.46\textwidth]{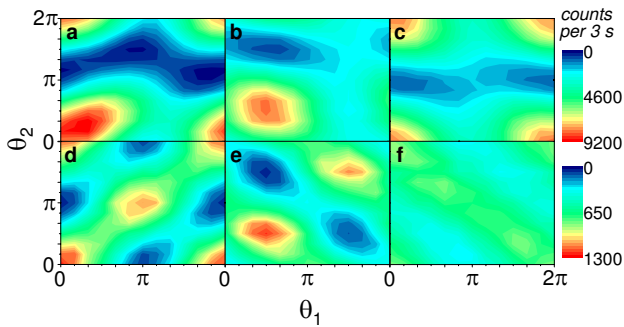}
\caption{{ Two-photon coincidence pattern.}
Coincidence counts of $\langle n_1 n_2 \rangle$
in a subspace $(\theta_1, \theta_2)$ of the parameters of local unitary operation.
The subspace is chosen to contain the positions of two of the local extremum values
of the two-qubit correlation
such that
$\lambda_l$, $- \lambda_{l}$, $\lambda_{l'}$, and $-\lambda_{l'}$ are located
at $(\theta_1, \theta_2) = (0,0)$,
$(\pi, \pi)$, $(\pi/2, \pi/2)$,
and $(3 \pi / 2, 3 \pi / 2)$, respectively:
{(a),(d)} $l = 1$ and $l' = 2$,
{(b),(e)} $l = 3$ and $l' = 1$,
{(c),(f)} $l = 2$ and $l' = 3$.
Three panels {(a)-(c)} are for step $k=0$, while {(d)-(f)} for $k=3$.
At step $k=3$ with maximal distillation,
$\langle n_1 n_2 \rangle$ has the same information
as the two-qubit correlation.
Panels {(d)-(f)} show a simple symmetric pattern, clearly distinct from {(a)-(c)}, and reveal
the quantum correlation in $\rho$ masked by classical correlation of local qubit information.
}
\label{scan}
\end{figure}

We discuss our experimental results.
DOPs 1 and 2 approach zero for larger $k$; see Fig.~\ref{iteration}.
They vanish below $\lesssim 0.1$ already at $k=3,4$, implying that the
maximal distillation is almost done.
This conclusion is supported by the fact~\cite{Lee}
that the error caused by partial distillation with DOPs $\lesssim 0.1$
is of the order of $(\mathrm{DOP})^2$
in the determination of $C(\rho)$ by $W_k$. 
It is also supported by the pattern of $\langle n_1 n_2 \rangle$
around the local extrema $\lambda$'s [see Fig.~\ref{scan}]:
At $k=0$, classical correlation of local qubit information leads to a complicated pattern,
while at $k=3$, it is removed, resulting in
a simple pattern.

At each step $k$, we measure $\lambda$'s and $s_0$,
and plot $\textrm{Tr} (W_k \rho) \equiv (1-\lambda_1-\lambda_2+q \lambda_3)/s_0$
and $\textrm{Tr} (W_{\textrm{dis},k} \rho_{\textrm{dis},k}) \equiv \textrm{Tr} (W_k \rho)/s_0$
in Fig.~\ref{iteration}(b).
We show that
$-2 \textrm{Tr} (W_k \rho)$
provides a lower bound of $\mathcal{C}(\rho)$~\cite{SUPP}.
The lower bound is the largest among those obtained from the witnesses constructed
with varying local unitary operation under the filtering setting fixed at step $k$.
Namely, $W_k$ and $W_{\textrm{dis},k}$ are the best witnesses at step $k$.
At the maximal distillation step, $W_k = W_\rho$.
For comparison,
$\mathcal{C}$'s are estimated~\cite{Wootters_conc} by
QST and maximum likelihood estimation at each $k$.
The distillation enhances concurrence
from $\mathcal{C}(\rho) \simeq 0.18$ to $\mathcal{C}(\rho_\textrm{dis}) \simeq 0.33$,
and
$-2 \textrm{Tr} (W_k \rho)$ approaches to the QST-concurrence $\mathcal{C}(\rho)$ 
within $0.03$ at $k=3,4$.
Given experimental uncertainty (see below),
the optimal witnesses $W_\rho$
and $W_{\rho_\textrm{dis}}$ are constructed at $k=3,4$,
and directly provide
$\mathcal{C}(\rho)$ and
$\mathcal{C}(\rho_\textrm{dis})$.

We mention the imperfections in the data,
(i) the incomplete erasure ($\lesssim$ 0.1) of DOPs,
(ii) the difference ($\lesssim 0.03$) between the concurrence
by $W_{k=3,4}$ and that by QST,
and (iii) the fluctuation ($\lesssim 0.03$)
of $\mathcal{C}(\rho)$ by QST
at different $k$'s.
They might come from
unwanted group delay~\cite{HKim} ($<$ 10 $\mu$m, depending on the filter parameters)
between the polarization components in the local filters
and/or
random fluctuation of coincidence counts due to the Poissonian statistics;
other error sources may exist.
Imperfection (i) might enhance (ii), as
it leads small error of $\sim (\mathrm{DOP})^2$ in the determination of $\mathcal{C}(\rho)$ by $W_\rho$;
this error is not included in the error bars in Fig.~\ref{iteration}(b).
Given these uncertainties,
it is reasonable to stop the distillation at the step ($k=3$) of
DOPs $\lesssim 0.1$.
Our Monte Carlo simulation reproduces qualitatively the same features.

We discuss the efficiency of our scheme.
At $k=3,4$, the error bar of $\mathcal{C}(\rho_\textrm{dis})$
determined by $W_{\rho_\textrm{dis}}$ is about three times smaller than
that ($\sim \pm$ 0.02) by QST.
This is obtained with
the time cost of 520 seconds in our scheme
(240 for three distillation steps, 160 for determining the settings for $\lambda$'s,
120 for measuring $\lambda$'s) and 160 seconds in QST
(only for quantification of $\rho_\textrm{dis}$).
Our scheme is more efficient than QST, as its error bar
will be roughly two times smaller than that of QST with 520 seconds.
Our simulation indeed shows~\cite{SUPP} that for the distillation and quantification of $\rho_\textrm{dis}$,
QST requires a number of measurements roughly two times larger than our scheme.
The better efficiency comes from the fact that
in our scheme the relevant quantities ($\lambda$'s, which {\em linearly} depend on measured counts)
are directly accessed,
using $W_\rho$ and the SLOCC invariance of $\mathcal{C}$,
while in QST the estimation of $\mathcal{C}$ requires
redundant information such as local properties of $\rho$
and is
sensitive to the estimation error due to the nonlinear dependence of $\mathcal{C}$ on $\rho$.

Our scheme is extendible to Greenberger-Horne-Zeilinger (GHZ) entanglement~\cite{GHZ}; see Ref.~\cite{SUPP}. The iterative distillation works for $N$-qubit (pure or mixed) states~\cite{Verstraete_normal}. It is useful for detection of `hidden' non-locality~\cite{Kwiat} of multiqubits. The quantification is extendible to three-qubit GHZ entanglement; it does not work for W-state entanglement, which behaves differently from GHZ under SLOCC. For a pure state, the optimal witness is obtained from a GHZ witness $W_\textrm{GHZ} = 3I/4 - |\textrm{GHZ} \rangle \langle \textrm{GHZ}|$ by applying the SLOCC operation for maximal distillation, then the exact value of a GHZ entanglement measure $\mathcal{T}_3$~\cite{SUPP} is obtained without QST.
For mixed states, the same procedure gives a reasonably good lower bound of $\mathcal{T}_3$.
This quantification is powerful, as it is a
formidable task to compute a GHZ entanglement measure 
for mixed states even with the QST information.

We have shown how to efficiently identify unknown two-qubit entanglement.
It is the first time, to our knowledge, that the optimal witness has been constructed
for entanglement quantification without QST.

We thank J. Eisert, P. Kwiat, L. C. Kwek, Tzu-Chieh Wei, and H. Weinfurter for discussions,
NRF (Mid-career Researcher, 2009-2332),
and KRISS project (``Single-Quantum-Based Metrology in Nanoscale'').
H.S.P. and S.S.B.L.
equally contributed to this work.

\end{document}